Title
# Ejecta distribution from impact craters on Ryugu: possible origin of the bluer units

Authors
**Naoyuki Hirata** [a, *], **and Ren Ikeya** [a]

* Corresponding Author E-mail address: hirata@tiger.kobe-u.ac.jp

**Authors' affiliation**
[a] Graduate School of Science, Kobe University, Kobe, Japan.

**Proposed Running Head:** Ejecta on Ryugu
Editorial Correspondence to:
Dr. Naoyuki Hirata
Kobe University, Rokkodai 1-1 657-0013
Tel/Fax +81-7-8803-6566

**Highlights**
- We calculated the distribution of ejecta blankets from actual craters on Ryugu
- Ejecta emplacement could explain the bluish color of the equatorial ridge
- Ejecta emplacement does not fully explain the bluish color of Tokoyo Fossa

Abstract

Asteroid (162173) Ryugu was the first spinning-top-shaped asteroid to be closely approached by a probe, the Hayabusa2 spacecraft, which sent numerous high-resolution images of Ryugu to the Earth and revealed the nature of this type of asteroid. One of the notable features of Ryugu is the equatorial ridge, which is considered the result of rapid spin in the past. Despite the advanced age of the ridge, indicated by the presence of numerous craters, the ridge exhibits a bluish color, indicating that it is covered with fresh material. In addition to Ryugu, many other asteroids have similar blue areas, which are considered the result of ejecta emplacement. We examined the distribution of ejecta blankets from actual craters on Ryugu to assess ejecta emplacement as a possible origin of Ryugu's bluer units. We determined that when Ryugu's rotation was fast, ejecta from craters formed at lower latitudes accumulated along the equator, which may explain the bluish color of the equatorial ridge. On the other hand, ejecta emplacement does not fully explain the bluish color of Tokoyo Fossa, although we attempted to find the corresponding ejecta blankets.

# 1. Introduction

Ryugu is a spinning top-shaped body, with a mean radius of 448.2 m and an equatorial ridge with a mean altitude of 502.5 m (Watanabe et al., 2019). The equatorial ridge is a result of a period of rapid spin in the past; the geopotential and slope analysis of the ridge indicates that Ryugu's rotation period (T) was once 3.5 to 5 h, although it is now 7.627 h (Watanabe et al., 2019). Ryugu's equatorial ridge has numerous craters, implying that it is a fossil structure (Hirata et al., 2020). The rotation period of small asteroids, including Ryugu, varies with time as a result of the so-called YORP effect (e.g., Walsh et al., 2008).

Ryugu's surface features a red-blue color dichotomy (Fig. 1), with (1) bluer materials distributed at the equatorial ridge, Tokoyo Fossa (an elongated depression in the southern hemisphere), and polar regions and (2) redder materials distributed in the mid-latitude regions (Sugita et al. 2019). Imagery and an artificial crater experiment by Hayabusa2 (Morota et al. 2020; Arakawa et al. 2020) show that (1) freshly exposed faces of boulders have a bluer color; (2) after touchdowns by the spacecraft, reddish dust was spread around the touchdown or impact sites; (3) after the artificial cratering experiment, reddish faces of boulders, such as the turtle rock, were altered to a bluer color; and (4) some small, fresh craters exhibit blue crater floors (for example, craters #5, #10, #16, #21, #31, #34, #41, #43, #47, #57, #58, and #61; crater numbering corresponds to Hirata et al. (2020) and Noguchi et al. (2021)). Morota et al. (2020) interpreted that (1) the bluer materials are relatively fresh, and become more reddish through surface metamorphic processes, (2) the surface is covered with a few meters of the redder materials while the interior is composed of bluer materials, (3) solar heating plays a major role in the surface metamorphic processes, (4) the bluer materials at the polar regions can be attributed to lower solar illumination in these regions, (5) the bluer materials at the equatorial ridge were likely exposed through mass wasting, from the equator (which is topographically highest) to mid-latitudes (topographically lowest), and (6) the origin of the bluer unit on Tokoyo Fossa was still unexplained.

Some other small bodies in the solar system also display a dichotomy of redder and bluer areas. For example, on the Martian moon, Phobos, bluer regions are distributed around or east of the Stickney crater. These regions are referred to as the blue units (Bibring et al. 1989; Ksanfomality et al.

1990; Avanesov et al. 1991; Patsyn et al. 2012; Pieters et al. 2014). Asteroid (243) Ida also exhibits redder and bluer spectral units (Veverka et al., 1996; Carlson et al. 1994). On asteroid (2867) Šteins, areas around the Diamond crater are slightly bluer than the rest of the surface (Schröder et al. 2010). Although the origin of these bluer areas is still debated, their association with ejecta emplacement has been proposed. Thomas (1998) proposed that ejecta blankets from the Stickney crater may correspond to the blue units of Phobos' surface if the rotational period of Phobos was slightly longer in the past. Kikuchi (2021) proposed that a combination of ejecta blankets from both the Stickney and Limtoc craters could explain the distribution of the blue units more adequately. Geissler et al. (1996) proposed that ejecta blankets from the Azzura crater correspond to the blue units of Ida. These studies calculated trajectories of ejecta from craters of various objects and examined whether the estimated landing locations of the ejecta corresponded to the distribution of blue units over the object's surface. Likewise, we calculated trajectories of ejecta from Ryugu's craters and determined whether their estimated landing locations correspond to Ryugu's blue units.

## 2. Analysis

Assuming various rotation periods for Ryugu, we numerically calculated the spread of the ejecta blankets. The methodology is similar to the authors' previous work, Hirata et al. (2021), but uses Ryugu's actual shape model, produced by Watanabe et al. (2019).

### 2.1. Initial launch position, velocity, and volume

First, we needed to define a horizontal plane to set the initial launch position and velocity of a particle ejected from an impact crater, because the highly irregular shape of the asteroid does not implicitly define a horizontal plane for the crater. Three points (A, B, and C) were selected on the surface of the shape model such that (1) the points were 1.2 crater rim radii away from the crater center, and (2) the points were evenly spaced around the crater center (at angles of 120° from the crater center) (Fig. 2a). We used 1.2 crater rim radii to define a horizontal plane to avoid undulations of the crater rim and floor. The horizontal plane was then defined as a plane passing through the three points A, B, and C. The center and size of the Ryugu craters were obtained from the crater database of Hirata et al. (2020).

On the other hand, some of craters overlap each other, and the existence of an adjacent crater may affect our definition of the horizontal plane: for example, a pair of crater #10 and #4. In order to evaluate this, we made Appendix A, where we selected points ABC of crater #10 so that the points do not overlap with crater #4 and calculated the ejecta distribution of crater #10.

The initial launch position, velocity, and volume of ejecta launched from each crater were determined using the scaling law developed by Housen and Holsapple (2011). Although they provide eight sets (C1–C8) of scaling parameters for a variety of materials (see also Table B1 in Appendix B), we basically used the scaling parameters of C4: dry sand targets in the gravity regime in this study. This is reasonable because Arakawa et al. (2020) suggested that the surface of Ryugu is composed of cohesionless sand-like material and its cratering occurs in the gravity-dominated regime, based on an artificial impact crater experiment. Nonetheless, the material properties of Ryugu are poorly known. Thus, in order to evaluate the difference in sets of scaling parameters, we made Appendix B.

We assume that the initial launch velocity and volume of the ejecta were axially symmetric. Although the motion of a particle in the excavation flow might be affected by the Coriolis force or local topography, we ignored such effects. Based on this assumption, the initial launch velocity ($v_{ej}$) of a particle in gravity regime is expressed as

$$v_{ej} = C_1 \left(H_1 \sqrt[3]{\frac{4\pi}{3}}\right)^{-\frac{2+\mu}{2\mu}} \sqrt{gR_c} \left(\frac{x}{R_c}\right)^{-\frac{1}{\mu}} \left(1 - \frac{x}{n_2 R_c}\right)^p, \quad (n_1 a \leq x \leq n_2 R_c)$$

(1)

where $x$ is the distance between the launch position and impact point, $g$ is the surface gravity, $R_c$ is the apparent crater radius, $a$ is the projectile radius, and the rest are scaling parameters for C4 ($\mu = 0.41, C_1 = 0.55, H_1 = 0.59, n_1 = 1.2, n_2 = 1.3, p = 0.3$) defined by Housen and Holsapple (2011). We assumed that $n_1 a = 0$, because the projectile size is much smaller than the crater size. Note that $n_2 R_c$ represents the crater rim radius ($R_c$ is not the crater rim radius). The point at which a particle crosses the horizontal plane defines the initial launch position ($x$) and velocity ($v_{ej}$) of the particle (Fig. 2B). If the position was below Ryugu's surface, it was adjusted along the vertical direction to the surface of the shape model. Here, $x$ and $R_c$ are defined as the distances from the crater center on the horizontal plane. The launch angle of the ejecta is 45° from the horizontal plane in any direction of $x$ (Housen and Holsapple, 2011).

The total volume ejected from inside $x$ (i.e., ejecta volume with an initial velocity greater than $v_{ej}(x)$) is given by the following equation from Housen and Holsapple (2011):

$$V = kx^3, \quad (0 \leq x \leq n_2 R_c) \quad (2)$$

where $k = 0.3$ is a constant for C4 (see also Table B1 in Appendix B). To discretize Eq. (2), we assumed that the ejecta volume at launch was axially symmetric. The volume of particles launched from the polar coordinate $(x_i, \theta_j)$ is given by the ejecta volume launched from a small area bounded by a radial distance between $x_{i+1}$ and $x_i$, and a small angular range of $\Delta\theta$:

$$V_{i,j} = k\frac{\Delta\theta}{2\pi}[x_i^3 - (x_i - \Delta x)^3] \quad , (3)$$

where

$$x_i = i\Delta x, \quad \Delta x = \frac{n_2 R_c}{N_1}, \quad \Delta\theta = \frac{2\pi}{N_2}, i = \{1, \ldots, N_1\}, j = \{1, \ldots, N_2\} \quad . (4)$$

In this study, we set $N_1 = 1000, N_2 = 360$.

## 2.2. Trajectories of ejecta

When the asteroid rotates with an angular velocity given by the vector $\mathbf{\Omega} = \{0, 0, \omega\}$, the equation of motion in the asteroid-fixed frame is given by (Scheeres et al., 2002):

$$\ddot{\mathbf{r}} + 2\mathbf{\Omega} \times \dot{\mathbf{r}} + \mathbf{\Omega} \times \mathbf{\Omega} \times \mathbf{r} = \mathbf{a}_g \quad , (5)$$

where $\mathbf{r}$ is the position vector of an ejecta particle relative to the asteroid center, and $\mathbf{a}_g$ is the gravitational acceleration acting on the particle. The origin of the coordinate system is the center of the asteroid, the z-axis is taken as the rotational axis, and the xy plane is taken as the equatorial plane of Ryugu. When the ejecta particle is more than three asteroid radii from the asteroid center ($|\mathbf{r}| > 3R_a$), $\mathbf{a}_g$ is obtained from a single point mass, equal to the asteroid mass:

$$\mathbf{a}_g = -\frac{GM}{|\mathbf{r}|^3}\mathbf{r} \quad , (6)$$

where the asteroid mass is $M = 4.5 \times 10^{11}$ kg and one asteroid radius is $R_a = 448.2$ m based on Watanabe et al. (2019). When the ejecta particle is within three asteroid radii of the asteroid center ($|\mathbf{r}| < 3R_a$), $\mathbf{a}_g$ is obtained from a point cloud placed at intervals of 5 m, expressing the irregular shape of Ryugu (Fig. 2C). By assuming that Ryugu is homogenous in density, each point of the point cloud represents an individual point mass, and $\mathbf{a_g}$ can then be obtained from the sum of the gravitational accelerations from each point mass. We used the so-called bucket method to improve the efficiency of the calculation. This method involved dividing the space around Ryugu to 100-m-spaced cubic partitions (or buckets). If a bucket was more than 100 m from the ejecta particle, the 5 m spacing points within the bucket were collectively calculated using the calculated geometric center and the total mass of the bucket (Fig. 2C).

Given the initial launch position and velocity of the particle, as described by Eq. (1), the trajectory of the particle was obtained by numerically integrating Eq. (5) until one of three outcomes occurred: (1) the particle reached an altitude greater than 168 asteroid radii (i.e., $|\mathbf{r}| > 168 R_a$); (2) the particle was below the asteroid surface (i.e., inside of the shape model); or (3) the particle had revolved around the asteroid more than five times. Note that the threshold of 168 asteroid radii corresponds to Ryugu's Hill radius. The trajectories were calculated in 1 s steps. When the particle landed on the asteroid surface, the location (latitude and longitude) of the particle was recorded. The ejecta thickness at a point was then obtained from the sum of the volume of particles falling within a 1° colatitude radius circle around the point. We did not consider lateral movements after landing, such as subsequent downslope motion or secondary ejection.

## 3. Results

Although we examined the ejecta distribution of all 77 craters on Ryugu, we present only four craters here: (#5, #16, #21, and #31). Note that the four craters are clear examples of bluish crater floors.

The ejecta distribution of crater #5 (50.56°S, 9.84°E, 173 m in diameter) is shown in Fig. 3 (left). This is the largest blue crater on Ryugu and the largest crater in the southern hemisphere. When T = 10000 h, (1) ejecta distribution was mostly symmetric around the crater, with slightly more ejecta accumulated on the southern slope of the equatorial ridge than on the northern slope, (2) ejecta did not accumulate on the antipodal point of the crater, (3) 95.2% of ejecta launched from the crater accumulated on the surface of Ryugu and the rest escaped, (4) the majority (97.9%) of ejecta landing on Ryugu's surface accumulated on the southern hemisphere, and (5) ejecta thickness around the crater rim was ~1–10 m. When T = 3.5 h or T = 7.627 h, a bow-shaped front line (along SE to NW direction) appeared to the east of the crater, and ejecta thickness to the east of the line was very thin. The generation of this folded ejecta blanket was reported by Dobrovolskis and Burns (1980) and Hirata et al. (2021). When T = 3.5 h, (1) ejecta mostly accumulated on the west of the crater, (2) the ejecta pattern appeared as a faint narrow band along the opposite latitude (50.5°N), with a thickness of approximately 0.01 m, (3) ejecta tended to accumulate more on the east-facing side of boulders than on the west-facing side, (4) ejecta tended to accumulate more on the south-facing (north-facing) side of boulders that are to the north (south) of the crater than on the north-facing (south-facing) side, (5) 94.9% of ejecta launched from the crater accumulated on the asteroid surface, while the rest escaped, and (6) 96.1% of ejecta landing on Ryugu's surface accumulated on the southern hemisphere.

The ejecta distribution of crater #16 (6.01°N, 308.75°E, 77.1 m), one of the craters on the equatorial ridge, is shown in Fig. 3. When T = 10000 h, the ejecta distribution was mostly symmetric around from the crater. On the other hand, when T = 3.5 h, most ejecta accumulated along the equatorial ridge, with a thickness in the order of 0.01 to 1 m. Ejecta particles that land along the equator are ones that land near crater #31 when the rotation was slow. In addition, ejecta tended to accumulate more on the east-facing side of the boulders than on the west-facing side. When T = 10000 h, 80.3% of ejecta launched from the crater accumulated on the surface and 95.2% of ejecta landing on Ryugu's surface accumulated at latitudes < 20° (N or S). When T = 3.5 h, 77.3% of ejecta launched from the crater accumulated on the surface, while 98.0% of ejecta landing on Ryugu's surface accumulated in the lower latitudes (< 20° N or S). When T = 3.5 h or 3.0 h, ejecta hardly accumulated in large crater depressions, such as Urashima. The ejecta distribution was similar for #21 (8.44°S, 119.53°E, 69.0 m) and #31 (17.20°S, 307.19°E, 48.8 m), the craters near the equatorial ridge (Fig. 4).

Overall, the ejecta blankets tend to blow toward the west when the rotation period is short. Ejecta travel westward, due to the Coriolis force, and therefore tend to collide with the east-facing, but not the west-facing, sides of boulders. Ejecta from craters near the equator accumulated near the crater rim when the rotation period was long, and spread to the entire equator when the rotation period was short.

## 4. Discussion

Evidence from craters #16, #21, and #31 indicates that with a shorter rotation period, ejecta from craters formed at lower latitudes accumulate along the equator. Therefore, ejecta emplacement could explain the origin of the blue units on the equatorial ridge. These blue units may have been either fed by multiple craters over Ryugu's history or formed by a single crater. Moreover, the accumulation of ejecta on the equator may explain the formation of the equatorial ridge (Ikeya and Hirata, submitted). Asteroid (101955) Bennu also has a slightly bluer equatorial ridge (DellaGiustina et al. 2020), which may be a result of ejecta emplacements, based on our calculations.

Our simulation implies that when the rotation period is short, ejecta fly toward the west and preferentially accumulate on the east-facing side of mounds (such as boulders), but barely accumulate in depressions (such as craters) and on the west-facing side of mounds. This is consistent with the reddish floors of Ryugu's large craters (Urashima, Kolobok, and Momotaro craters). On the other hand, we could not find clear examples of boulders with an east-west color dichotomy; for example, the Ejima boulder (Fig. 1) does not show an east-west color dichotomy.

Ejecta emplacement does not sufficiently explain the origin of the bluer units on Tokoyo Fossa. We attempted to find ejecta blankets corresponding to the bluer units on Tokoyo Fossa and examined all actual craters assuming various rotation periods, however, we did not find any clear examples. The origin of the bluer unit at Tokoyo Fossa may be associated with the formation of the so-called western bulge, rather than ejecta emplacement. The western bulge is proposed to have been responsible for the deformation of Ryugu, via a process that occurred on only the western side of the asteroid during a short rotational period of Ryugu in the past, while the rest of Ryugu was left structurally intact (Hirabayashi et al. 2019). Also, Scheeres (2015) proposed that regolith landslides occur on the surface of rapidly spinning asteroids, so the occurrence of such a landslide could be taken into consideration as an alternative explanation for the formation of the western bulge. Nonetheless, when T = 7.627 h (Fig. 4, left) ejecta from crater #21 preferentially accumulated on Tokoyo Fossa and the east of the crater, which could explain the bluer units there.

Most of the craters with bluer floors are concentrated in an area between 270°E and 360°E, in the southern hemisphere. This may be attributed to the ejecta blankets of crater #5 (or crater #10, Kibidango). Crater #5 is the largest blue crater on Ryugu, and our numerical simulation indicates that ejecta blankets from it should have been deposited in the region of 270°E to 360°E, in the southern hemisphere, with a thickness of 0.1~1 m. It is possible that relatively fresh, bluish deposits originating from crater #5 do exist in the subsurface of this region, and that the overlying reddish layer is thinner here than the rest of Ryugu. If so, bluer craters should be more easily formed in this area, than on the rest of Ryugu.

Lastly, it should be noted that it is difficult to determine conclusively whether ejecta are more blue or more red. Ryugu's surface is covered with redder materials to a depth of a few meters, while the interior is composed of bluer materials (Morota et al. 2020), and therefore, the ejecta from small craters excavating less than a few meters in depth should be redder. In fact, ejecta from the artificial crater showed a redder color (Arakawa et al. 2020). The area around crater #47 (34.51°S, 263.82°E, 34.4 m in diameter) is covered in redder units and only the crater floor of #47 is bluer, although we expect that ejecta emplacement from #47 should have occurred in the area. A possible explanation for this is that the impact completely excavated the redder surface materials, but barely excavated any interior bluer materials. In fact, the depth of crater #47 is 3.6 m, relative to the surrounding topography (Noguchi et al. 2021), which is nearly equal to the thickness of redder materials.

## 5. Conclusion

We examined the distribution of ejecta blankets from actual craters on Ryugu to assess the role of ejecta emplacement in the formation of Ryugu's bluer units. We found that when Ryugu's rotation period was short, ejecta from craters formed at lower latitudes accumulated along the equator, which could explain the blue units now found on the equatorial ridge. The blue units on the equatorial ridge may have been formed by a single large crater, or may have been fed by ejecta from multiple craters over Ryugu's history. On the other hand, ejecta emplacement does not sufficiently explain the origin of the bluer units on Tokoyo Fossa, although we attempted to find the corresponding ejecta blankets. Most of the craters having a bluer crater floor are concentrated in an area in the southern hemisphere, between 270°E and 360°E, which may be attributed to ejecta blankets from crater #5, whose ejecta blankets have a thickness of 0.1~1 m in this area.

**Appendix A.**

Some of craters overlap each other. The existence of an adjacent crater may affect our definition of the horizontal plane. For example, when we define the horizontal plane of crater #10, point A exists on the floor of crater #4. Thus, we obtain a new horizontal plane of crater #10 through points A'B'C', which were obtained by rotating ABC 60° with respect to the crater center (Fig. A1), so that A'B'C' do not overlap with the floor of crater #4, and recalculate the ejecta blanket of crater #10 using the new horizontal plane (Fig. A2). As shown in Figure A2, the ejecta thickness is slightly different in each definition, but we find that the difference is on the order of 1 %.

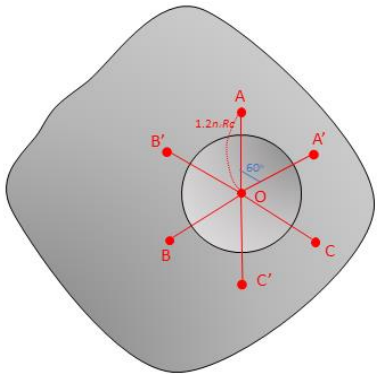

**Figure A1.** Difference of the definition of points ABC and A'B'C'.

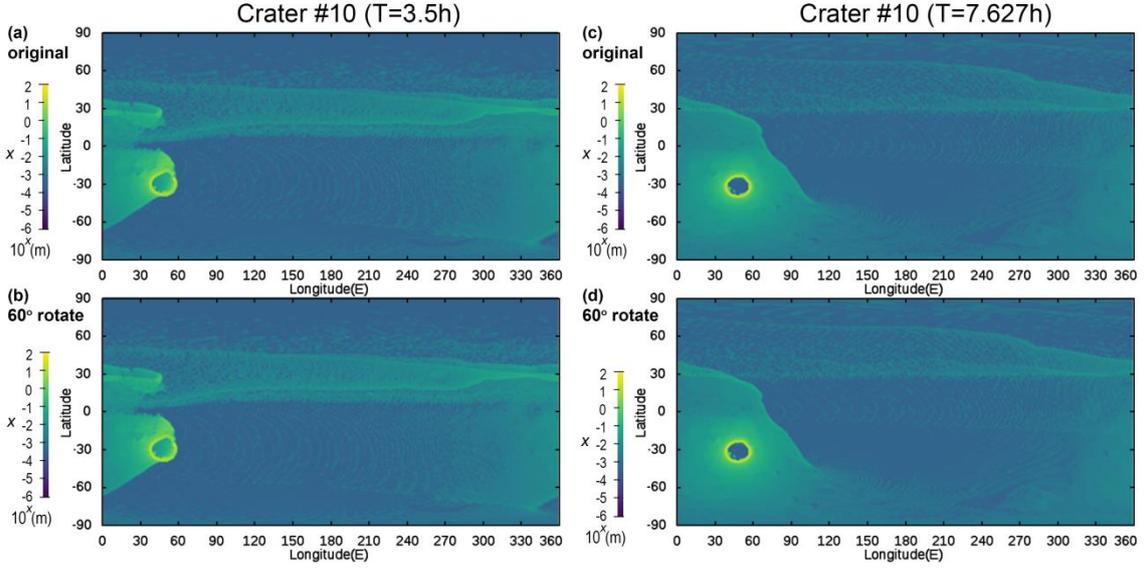

**Figure A2.** Ejecta distribution of crater #10 when T=3.5h (a, b) and T=7.627h (c, d) using the horizontal planes ABC (a, c) and A'B'C' (b, d).

### Appendix B.

We calculated the ejecta distribution assuming other sets of scaling parameters defined by Housen and Holsapple (2011). Housen and Holsapple (2011) provide 8 sets of scaling parameters which include four examples in gravity regime (C1, C4, C5, C6) and four examples in strength regime (C2, C3, C7, C8) (Table B1). The initial launch velocity ($v_{ej}$) of a particle in gravity regime is shown in Eq. (1). The initial launch velocity ($v_{ej}$) of a particle in strength regime is expressed as:

$$v_{ej} = C_1 \left(H_2 \sqrt[3]{\frac{4\pi}{3}}\right)^{-\frac{1}{\mu}} \sqrt{\frac{Y}{\rho_t}} \left(\frac{x}{R_c}\right)^{-\frac{1}{\mu}} \left(1 - \frac{x}{n_2 R_c}\right)^p, \quad (n_1 a \leq x \leq n_2 R_c) \quad \text{(B1)}$$

where $\rho_t$ and $Y$ are the target density and strength (Housen and Holsapple, 2011).

Figure B1 and B2 show ejecta distributions from crater #5 and #16 assuming T=3.5 h and scaling parameters for C1, C2, C3, C6, C7, and C8. Note that C5 is identical to C4. The case of C4 is shown in Fig 3. Figure B1 shows that, even in the case of other sets of scaling parameters, ejecta from crater #5 tend to accumulate an area in the southern hemisphere between 270°E and 360°E when Ryugu' rotation is fast. On the other hand, ejecta blankets in the area have a thickness of 0.1~1 m in the case of C1 to C6, while those have a thickness of 0.001~0.1 m in the case of C7 and C8. This is because most of ejecta in the case of C7 and C8 tend to accumulate around the crater rim. Figure B2 shows that ejecta from craters formed at lower latitudes accumulated along the equator. As a result, although the ejecta distribution is slightly different in each set, we consider that the difference of scaling parameters barely affects our conclusions.

Table B1. Scaring parameters used in ejecta model, based on Housen and Holsapple, (2011).

| No.* | C1 | C2 | C3 | C4 | C5 | C6 | C7 | C8 |
|---|---|---|---|---|---|---|---|---|
| Reg.** | G | S | S | G | G | G | S | S |
| $\mu$ | 0.55 | 0.55 | 0.46 | 0.41 | 0.41 | 0.45 | 0.40 | 0.35 |
| $k$ | 0.2 | 0.3 | 0.3 | 0.3 | 0.3 | 0.5 | 0.3 | 0.32 |
| $C_1$ | 1.5 | 1.50 | 0.18 | 0.55 | 0.55 | 1.00 | 0.55 | 0.60 |
| $H_1$ | 0.68 | - | - | 0.59 | 0.59 | 0.8 | - | - |
| $H_2$ | - | 1.1 | 0.38 | - | - | - | 0.40 | 0.81 |
| $n_2$ | 1.5 | 1.0 | 1.0 | 1.3 | 1.3 | 1.3 | 1.0 | 1.0 |
| $p$ | 0.5 | 0.5 | 0.3 | 0.3 | 0.3 | 0.3 | 0.3 | 0.2 |
| $\rho_t$ (kg/m3) | 1000 | 3000 | 2600 | 1600 | 1510 | 1500 | 1500 | 1200 |
| $Y$ (MPa) | - | 30 | 0.45 | - | - | - | $4 \times 10^{-3}$ | $2 \times 10^{-3}$ |

* Scaling parameters for Water (C1), Rock (C2), weakly cemented basalt (C3), sand (C4 and C5), glass micro-spheres (C6), sand/fly ash mixture (C7), and perlite/sand mixture (C8).

** The strength regime (S) or gravity regime (G).

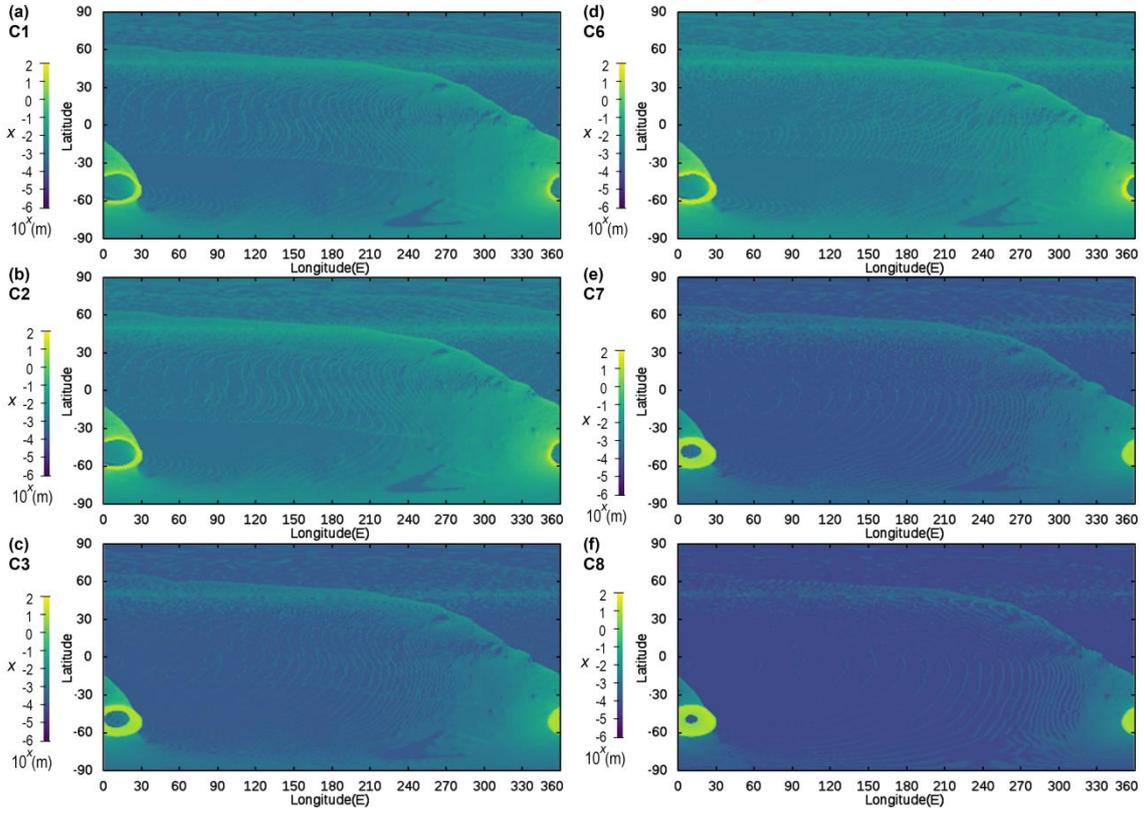

**Figure B1.** The ejecta distribution from crater #5 assuming T=3.5 h and scaling parameters for (a) C1, (b) C2, (c) C3, (d) C6, (e) C7, and (f) C8.

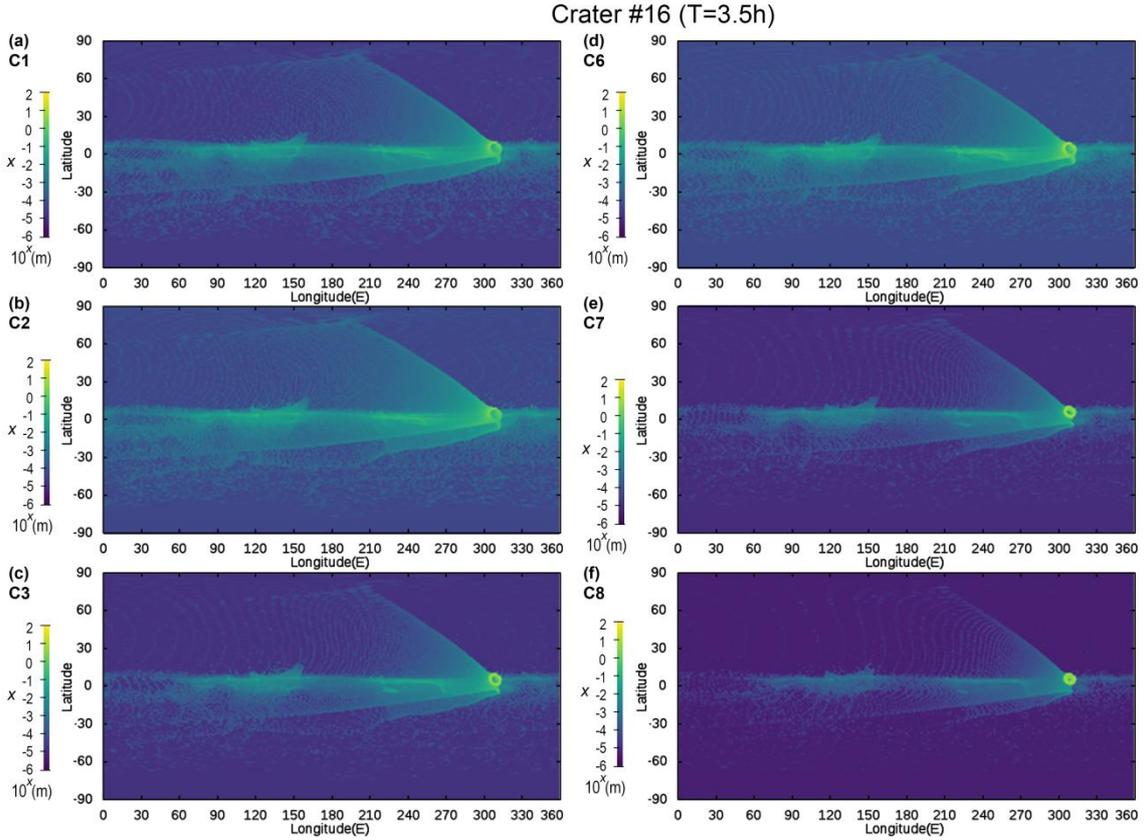

**Figure B2.** The ejecta distribution from crater #16 assuming T=3.5 h and scaling parameters for (a) C1, (b) C2, (c) C3, (d) C6, (e) C7, and (f) C8.

Acknowledgments

We thank all members of the Hayabusa2 mission team for their support in data acquisition. We appreciate two anonymous reviewers and their helpful comments. This work was partly supported by JSPS Grants-in-Aid for Scientific Research Nos. 20K14538 and 20H04614 and the Hyogo Science and Technology Association (NH). The shape model of Ryugu is freely available via the Data ARchives and Transmission System (DARTS) at ISAS/JAXA (https://data.darts.isas.jaxa.jp/pub/hayabusa2/paper/Watanabe_2019/README.html).

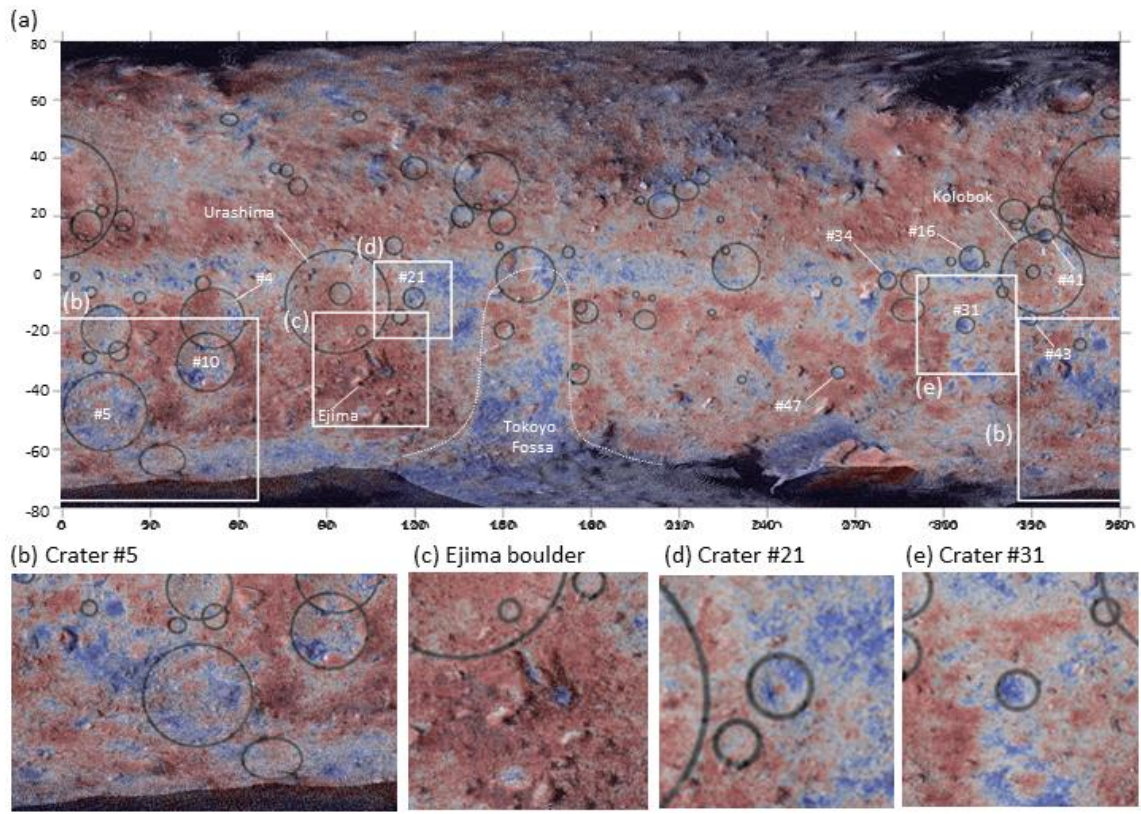

**Figure 1.** The distribution of redder and bluer units on Ryugu, adapted from Morota et al. (2020).

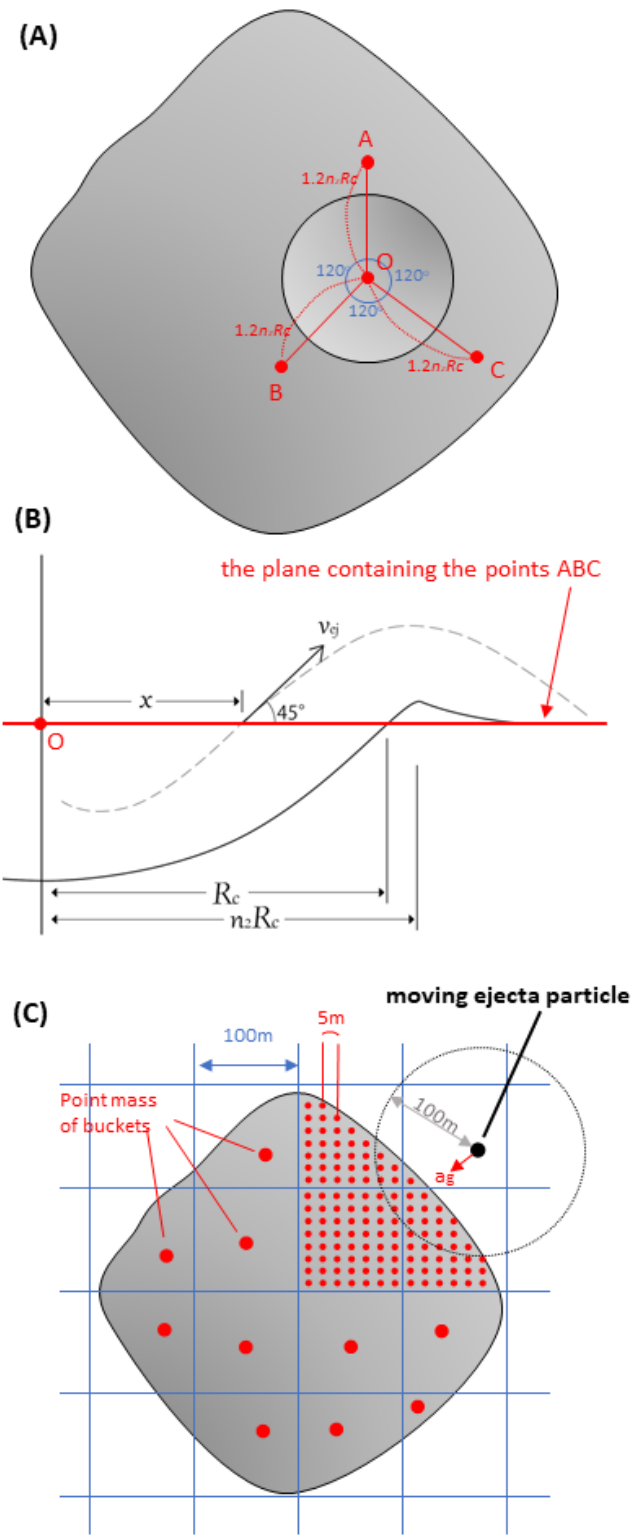

**Figure 2.** (a) Definition of the horizontal plane. (b) Definition of variables in Eq. (1) in Section 2.1 and the trajectory of a particle launched from location, $x$. (c) Method for obtaining the gravitational acceleration acting on a moving

ejecta particle.

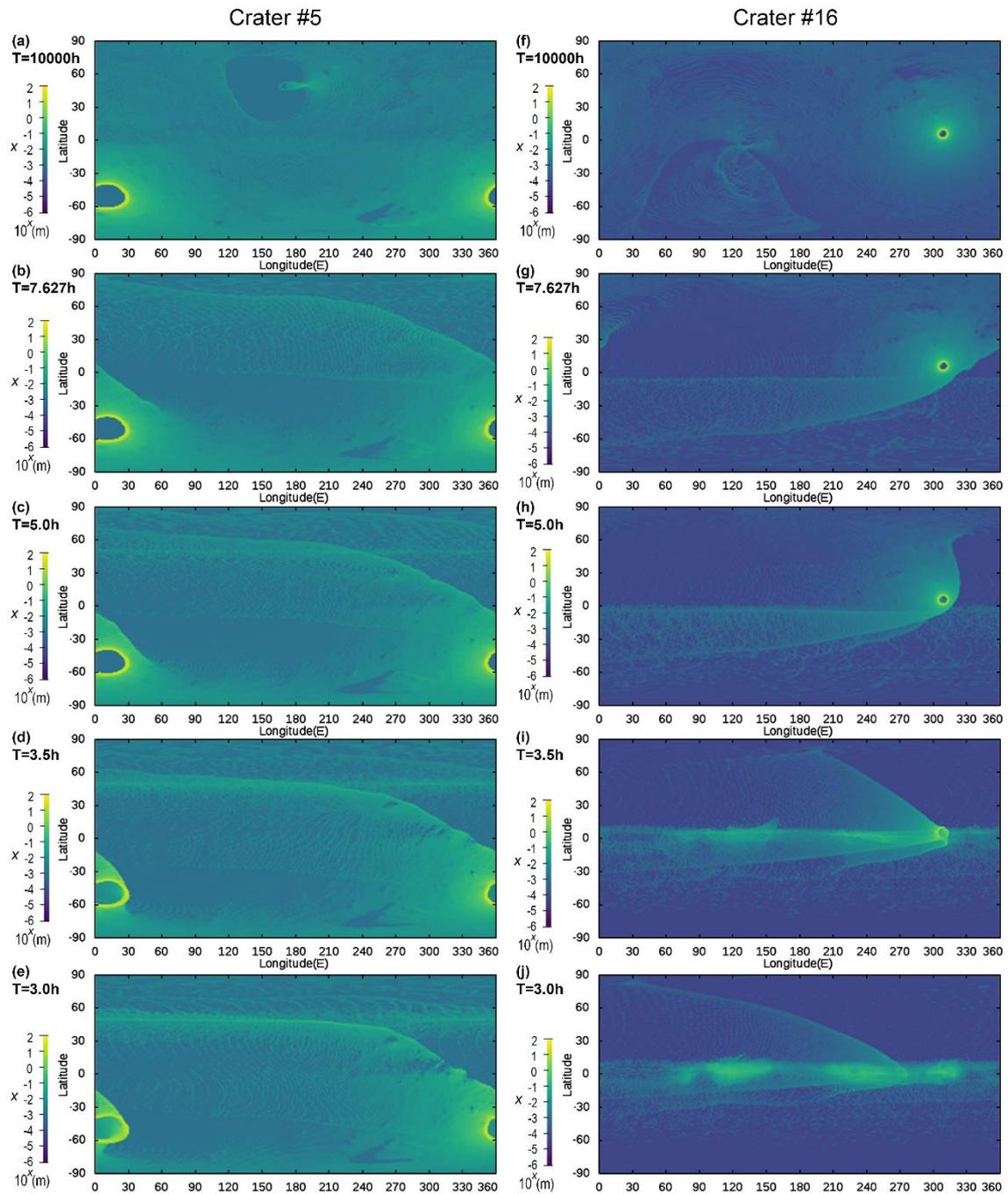

**Figure 3.** The global ejecta thickness from craters #5 (left) and #16 (right) when T = 10000, 7.627, 5.0, 3.5, and 3.0 h (from top to bottom). The color bar is presented in a logarithmic scale.

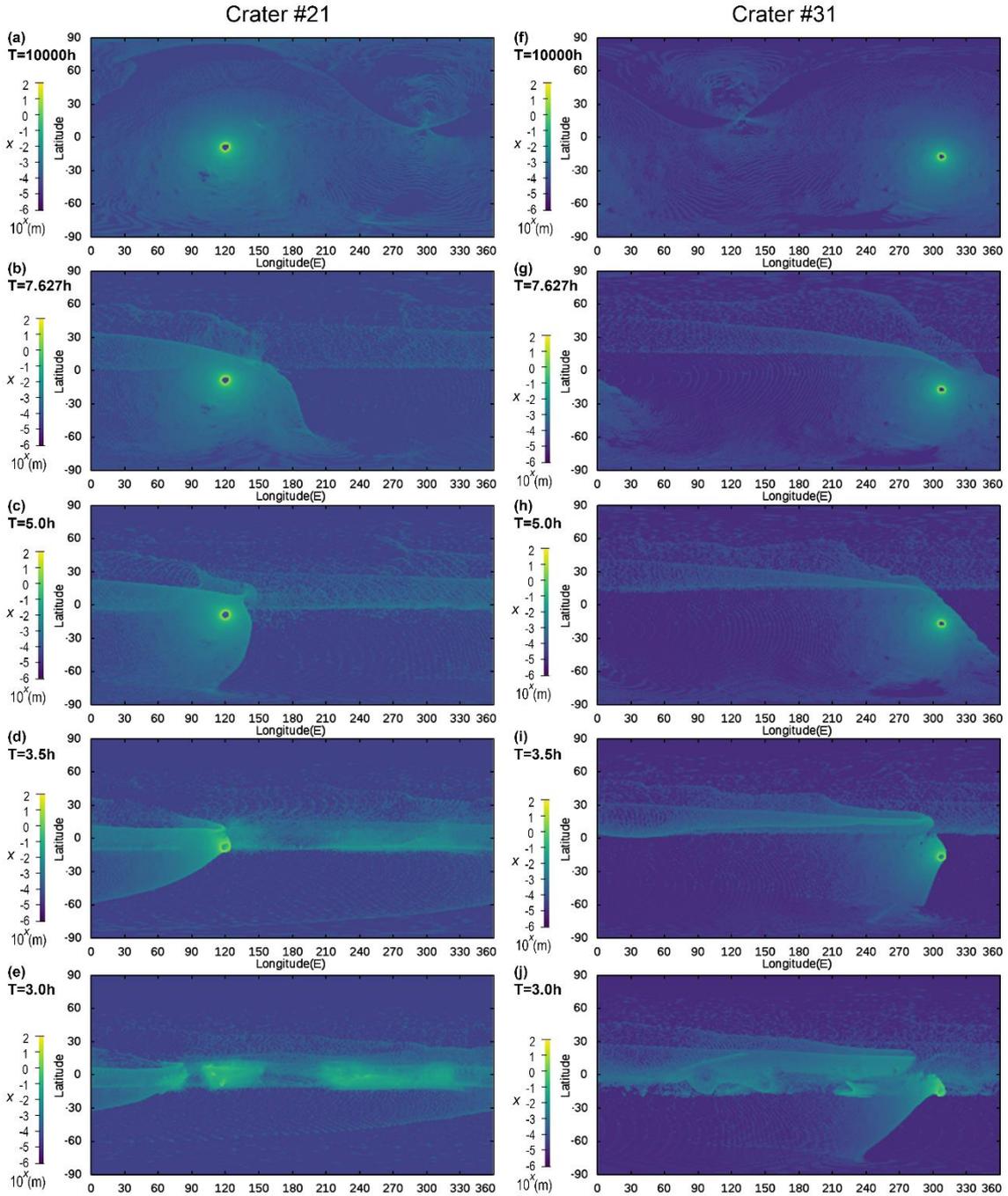

**Figure 4.** The global ejecta thickness from craters #21 (left) and #31 (right) when T = 10000, 7.627, 5.0, 3.5, and 3.0 h (from top to bottom). The color bar is presented in a logarithmic scale.

### References

Arakawa, M. et al., (2020), Artificial impact crater formed on the asteroid